\documentclass[12pt,a4paper]{article}

\pdfoutput=1
\usepackage{exscale,hyperref}

\newtheorem{conjecture}{Conjecture}
\newlength{\Bemerkung}

\begin{document}

\title{Extreme-value statistics of dimensions determining an observer's 
branch of the world?} 

\author{L. Polley}

\date{\small Institut f\"ur Physik, Oldenburg University, 26111 Oldenburg, Germany}

\maketitle

\begin{abstract}
In a many-worlds framework, combining decoherent histories and 
extreme-value statistics, it is conjectured that the (matrix) dimension of the Hamiltonian 
processing records and memories near the end of an observer's history is almost 
entirely located in a single branch of his/her wavefunction.  
\\[1mm]
Keywords: Many worlds, order statistics, backward causation.
%A many-worlds scenario of an observer's evolution is considered, in which decoherent 
%branches are governed by randomly chosen Hamiltonians, generating further branchings.
%In the huge ensemble of effective Hamiltonians near the observer's end, any chain reaction 
%(or other power-law distributed random process) among collected memories 
%would result in an extreme value of Hamiltonian's dimension which would locate 
%nearly all cognitive dimension in a single branch.         
\end{abstract}

\section{Introduction}

The theory of decoherence has been successful in explaining many of 
the classical features of an observer's world \cite{Zurek2003,SchlosshauerBuch2007}. 
Among the problems left unexplained is that of outcomes of a quantum measurement, which is
because decoherence does not require any collapse of a wave function%
\footnote{Using statistical operators in the context of decoherence is a matter of  
convenience; for an approach using state vectors only, cf.\ \cite{Zurek2007}.}, 
but is consistent with a many-worlds interpretation. The latter \cite{Everett1973}
is usually taken to imply that an observer's impression of living in a unique 
world does not reflect his/her true wave function, 
but rather reflects the mutual unobservability of decoherent branches, supposed to be
equivalent.  

The splitting into branches is a consequence of the Schr\"odinger equation applied 
to an act of observation, hence can be seen as a matter of first principles. But 
there is no such basis for assuming the equivalence of branches.
These may carry different weights in a {\em subjective\/} sense. 
``Massive redundancy can cause certain information to become objective, at the expense of 
other information'' as pointed out in \cite{Zurek2005}. However, while \cite{Zurek2005} 
proceeds by measuring redundance in information-theoretic terms, 
the notion of weight advocated here is rather that of \cite{Isham1994}, 
where decoherent histories of observations 
are defined by sequences of projection operators, and the weight assigned to a history is 
the dimension of the projected subspace. In the present paper, weight will be assigned in
proportion to the dimension of the observer's ``experiential subspace'', determined by an 
effective Hamiltonian of interacting memories and records {\em at the end\/} of a branch of 
history. 
The reason for assuming such a fundamental role of backward causation lies in 
order statistics \cite{David1981,Embrechts1997} of power-law ensembles, which comes into play 
here because the number of records and memories is conjectured to be pseudo-random and 
power-law distributed. Since the number of branches is largest towards their ends, the 
extremal draw, which turns out to be exhaustive, is most likely to occur there. 

The exact form of statistical ensemble conjectured is exponentials of power-law variates.
While power laws are ubiquitous empirically, with many scenarios for their generation 
conceivable \cite{ReedHughes2002,SimkinRoychowdhury2006}, exponentiation needs to be 
motivated. It is based on the observation \cite{Zurek2005} that records of 
quantum-mechanical measurements are, in normal circumstances, stored in subsystems provided as
fragments of the environment, 
$$
    {\cal E} = {\cal E}_A \otimes {\cal E}_B \otimes {\cal E}_C \otimes \cdots 
$$
For an observer to become aware of a record, he/she must physically interact with it.
This requires (as in all models considered in \cite{Zurek2005}) minimal dimensions
$$
    D_{{\cal E}_A} \geq 2 \qquad D_{{\cal E}_B} \geq 2 \qquad  D_{{\cal E}_C}\geq 2 \qquad
   \cdots
$$ 
If it is assumed that the records fluctuate in number, with power-law distribution
over the branches, it follows that the product of their dimensions is an exponentiated 
power-law variate.  

In this case, by a theorem of order statistics \cite{David1981,Embrechts1997},
the largest draw of the number of records exceeds the second-largest 
draw by a difference {\em increasing\/} with the size $N$ of the ensemble (number of branches) 
like some positive power $N^{1/\alpha}$. The exponentiated number (dimension) in the extremal 
branch, consequently, exhausts nearly all of the dimension summed over the branches. 

In section \ref{SecStatistics}, some extreme-value statistics is recalled.  
Section \ref{SecHeff} points out the role of effective Hamiltonians for the scenario envisioned.
Section \ref{SecExponentiation} elaborates on the exponentiation of power-law variates.  
Because of the branching structure, some care is needed to ensure statistically independent 
draws from a power-law ensemble, which is discussed in 
section \ref{SecFinalDraw}. 
The somewhat problematic role of amplitudes in the scenario is discussed in section 
\ref{SecAmplitudes}. Conclusions are given in section \ref{Conclusion}. 
Conjectures \ref{ConjHamiltonian}, \ref{ConjPowerLaw}, \ref{ConjFinalDraw} give the main ideas.

\section{Recalling some extreme-value statistics\label{SecStatistics}}

Order statistics \cite{David1981,Embrechts1997} provides the 
following theorem on outliers of power-law distributions. 
In an ensemble of size $N$ of random variables $X$ with a probability density 
proportional to  $X^{-1-\alpha}$ for large $X$, 
the difference of the largest and second-largest draw is 
(notation of \cite{Embrechts1997} corollary 4.2.13)
\begin{equation}   \label{FrechetSeparation}
    X_{1,N}-X_{2,N} = N^{1/\alpha} \, Y 
   \qquad \mbox{$Y$ = random variable independent of $N$} 
\end{equation}
The power law need not be exact, but the distribution must lie in the domain of attraction of 
the Fr\'echet distribution, which has a cumulative distribution function $\exp(-X^{-\alpha})$.
The point is that the difference in (\ref{FrechetSeparation}) becomes arbitrarily large 
with $N\to\infty$. By contrast, it would remain finite for normally and even log-normally 
distributed variables. 

In the present context, $N$ is the number of observer's branches, assumed to grow 
exponentially on average. Thus, if the observer's lifetime is $T$,
\begin{equation} \label{NumberBranches}
  N \propto e^{T/\Delta} 
\end{equation}
Equation (\ref{FrechetSeparation}) with $N\to\infty$ need not amount to any 
domination of the extreme for 
power-law ensembles themselves, since the second largest value would also be of order 
$N^{1/\alpha}$. However, if the variable is exponentiated like $2^X$, the extremal draw 
easily exceeds the summed-up variables of the rest of draws, since $2^{N^{1/\alpha}}\gg N$.

It should be noted that, when extremal values are taken over increasingly large 
ensembles, not every distribution lies in the domain of attraction of one the three 
possible limiting forms. 
{\em Exponentiated\/} power-law variates, with cumulative distribution function of the form
$F(Z) = \exp(-\beta (\ln Z)^{-\alpha})$, do not satisfy the convergence criterion 
(as given in corollary 3.3.8 of \cite{Embrechts1997}). 

The reason for proposing the fundamental role of backward causation in the scenario 
is that ``when the sum of [\ldots] independent heavy-tail random variables is large,
then it is very likely that only one summand is large'' \cite{Barbe2004}. Because of this, 
there is no advantage in alternative scenarios in which the extremal quantity is accumulated 
during an observer's lifetime. 
The extremal draw most probably occurs near the end of a branch, 
since the ensemble is largest there. Using (\ref{NumberBranches}) and estimating 
probabilities as proportional to instantaneous ensemble sizes, $p = e^{-n}$ for an
occurrence earlier than $T-n\Delta$.

A ``final draw'' mechanism for gathering the bits of experience (assuming  
power-law distribution) can be quite efficient. If bits were accumulated at a 
constant rate, their number would grow proportional to the observer's lifetime $T$.  
On the other hand, if the bits are assembled by supercritical, randomly terminated 
chain reaction substisting on a reservoir of uncollapsed superpositions towards the ends 
of an observer's branches, their number rises proportional to $N^{1/\alpha}$ 
\cite{ReedHughes2002,SimkinRoychowdhury2006} where $N$ is given by (\ref{NumberBranches}).
Thus, the number of bits grows exponentially with $T$.

\section{Random effective Hamiltonians\label{SecHeff}} 

In applications of quantum mechanics, one identifies an effective Hamiltonian relating to
the coordinates of interest. 
In the scenario considered here, each branching of the system-observer state is assumed 
to be accompanied by the emergence of a new effective Hamiltonian for each branch. 
All those Hamiltonians would, in principle, be determined by a large 
number of conditions contained in the exact initial state of the observer and his/her 
environment. For all practical purposes, the following should hold.   
\begin{conjecture} \label{ConjHamiltonian}
Initial conditions for an observer and his/her environment are sufficiently complex, such that
effective Hamiltonians, governing branches of the wavefunction, are pseudo-random. 
\end{conjecture}
The randomness is assumed to apply to an observer's processing of experiences as well.  
No collapse of a wavefunction is involved here. 

Von Neumann's scheme \cite{vNeumann1955} for a measurement described by a Schr\"odinger 
equation requires only slight augmentation. The joint evolution of system and 
recording device proceeds, accordingly, like
\begin{equation}  \label{vNeumannScheme}
   |\psi\rangle |\mathrm{blank}\rangle  ~ \longrightarrow ~ 
    \sum_{i=1}^n |i\rangle\langle i | \psi\rangle ~ |\mathrm{recorded}~i\rangle
\end{equation}
For simplicity, it is often assumed that the recorded states are uniquely
specified by the label given. This is what has to be changed for the present approach. 
As discussed in section \ref{SecExponentiation},
a readable record must lie in a subspace of two, or more, dynamically accessible dimensions. 
Von Neumann's scheme should therefore be augmented by postulating
\begin{equation}  \label{vNeumannAugmented}
     |\mathrm{recorded}~i\rangle \in {\cal H}_{\mathrm{recorded}~i} 
   \qquad\qquad \mathrm{dim}\,{\cal H}_{\mathrm{recorded}~i} \geq 2 
\end{equation}

\section{Exponentiating power-law variates \label{SecExponentiation}}

Rephrasing what has been stressed in \cite{Zurek2005} already, any bit of information 
that ``exists'' for a mind must participate in the process of cognition.
For this, the bit must be physically represented by a subsystem ready to reveal its 
existence by interaction, which requires dimension $\geq 2$.
For example, if some information would be stored in a particular neuron of the observer, 
it would have the dimensions of firing and resting, corresponding to the open and closed
state of an ion-pore molecule. Since this is the basis of neural activity, and since 
dimension is not lost when systems are coupled for collective action (the dimension 
being preserved in the number of different patterns that could be supported)
it would seem justified to assume a dimension of $2^X$ for the cognitive subspace in
which $X$ neurons are involved. Similarly, the information stored in a black grain 
of a photograph is available by absorption of light, which at least requires 
two energy levels.
In von Neumann's minimal scheme, where there is branching from a blank state to  
one-dimensional recorded states, acquisition of higher-dimensional systems is only 
postponed until the reading of the records. The presence of $|\mathrm{recorded}~i\rangle$ 
may be detected, for example, by some scattering experiment, which
would require $\geq 2$ outgoing channels.  
\begin{conjecture} \label{ConjPowerLaw}
The number of interacting records, each carrying dimension two at least, is power-law
distributed over the ends of an observer's branches. 
\end{conjecture}
On this basis, we can estimate the exhaustiveness of the extremal branch.
Let $X$ be the random number of bits, and $2^X$ the dimension of 
the observer's experiential subspace. 
If the observer ends up with $N$ branches, we can use (\ref{FrechetSeparation}) to 
estimate the fraction of dimension {\em outside} the leading branch to be smaller than
\begin{equation}   \label{nonleadingDimension}
    (N-1) \times 2^{X_{2,N}}\Big/ 2^{X_{1,N}} ~ = ~ 2^{-Z} \mbox{~~~where~~~}
  Z = N^{1/\alpha} + {\cal O}(\log N)
\end{equation}
since in every non-leading branch the number of bits is not greater than $X_{2,N}$. 

But is dimension an ``element of reality'', or rather a formal construct, invoked only in
order to enhance an otherwise insufficient effect? In applications of 
Fermi's Golden Rule, transition probabilities are proportional to the density of final 
states, thus indeed, proportional to the dimensions of subspaces.  
On the other hand, if subspaces emerge as products of qubits, it would be  
information-theoretic practice to count the bits, not the dimension. For large ensembles,
obviously, this makes a crucial difference in the statistical properties.

\section{Final draw of Hamiltonian\label{SecFinalDraw}}

Formula (\ref{FrechetSeparation}) applies to {\em independent random draws\/} with a 
power-law distribution of probabilities. It is not sufficient to have an ensemble in
which a quantity $k$ occurs in a number $k^{-\mu}$ of members.
For example, in randomly growing scale-free networks \cite{AlbertBarabasi2002}, a probability  
$P(k) \propto k^{-1-\alpha}$ develops for a node to have $k$ connections, if preferential 
attachment of new nodes to existing ones is assumed. But if the maximal node size has grown
to $k_\mathrm{max}$, the second-largest size is typically $k_\mathrm{max}-1$; there is no 
outlier emerging%
\footnote{In a classical network with preferential attachment superproportional to $k$,
the ``winner takes all'' \cite{AlbertBarabasi2002}. However, in the application here,
the winning cluster would grow during a large number of branchings, so most of it would 
be common to many branches.}.
Thus, it would be an {\em insufficient\/} scenario to regard an 
observer's branching tree as a network, and the mere termination of a branch as random draw, 
since the ends of branches are highly correlated by the common parts of their history. 
A similar problem occurs with power-law ensembles generated by Galton-Watson processes, 
like critical processes generating $P(k) \propto k^{-3/2}$ for the number of bits 
summed over all generations \cite{Pakes1971}.   

Statistically, some ``final draw'' near the end of each 
branch is required.  Observer's awareness here enters the scenario,
since in addition to a reservoir of records, awareness requires cognitive 
dynamics, supposedly (Conjecture \ref{ConjHamiltonian}) represented by some pseudo-random 
effective Hamiltonian. Matrix elements $H_{ij}$ would provide the links between clusters 
of records (including memories as neuronal records). Some kind of chain reaction could  
be envisioned, which would be the standard scenario for generating power-laws 
\cite{ReedHughes2002,SimkinRoychowdhury2006}.
It would have to be supercritical (exponential growth) and terminate with a 
constant incident rate (exponential distribution of runtime). If the number of generated 
bits grows like $e^{\beta t}$, and if the probability for runtimes decreases like 
$e^{-\gamma t}$, the probability for generating $k$ bits in the process is  
$ P(k) \propto k^{-1-\gamma/\beta}$. It should be stressed, however, that $H_\mathrm{eff}$
is supposed to {\em result\/} from the chain reaction, not to generate it. The generation 
would be accomplished, as suggested by Conjecture \ref{ConjHamiltonian}, by an
all-encompassing Hamiltonian%   
\footnote{Speculating in more detail, a ``final draw'' could work as follows.
Firstly, assume a hierarchy in the observer's experiences, with physiological events
at the top (determining ends of branches, guaranteeing for decoherence and statistical 
independence), followed by redundant observations of various degrees.
Secondly, allow for a redundant observation to be preserved in a state of superposition 
(uncollapsed) until, by one of the final Hamiltonians, one of its components becomes linked 
to the more 
basic experiences and their dynamical dimensions. That component, being part of a 
system-observer entanglement, will carry indications of the time at which the observation was 
originally made. It will thus fit in with the observer's subjective account of history.}. 
Conjecture \ref{ConjPowerLaw} with estimate (\ref{nonleadingDimension}) then suggests 
\begin{conjecture} \label{ConjFinalDraw}
Observer's one-world experience resides in (the matrix dimension of) the effective 
Hamiltonian processing records and memories near the end of his/her extremal branch.
\end{conjecture}

\section{Problematic role of amplitudes\label{SecAmplitudes}} 

A tacit assumption has been that the number of branches emerging in an observation are 
finite, and their probability amplitudes more or less equal. It should be noted that 
Born's rule can be derived on such a basis \cite{Deutsch1999}. On the other hand, 
a finite number of branches is an idealisation in which any remote possibilities, such as 
tunneling under a macroscopic potential barrier, are neglected. Clearly, such amplitudes are
extremely small, and in fact, their numerics is even more extreme than that of dimensional 
statistics. 
For example, if an observer can cope with 300 measurements per second, the amplitude in  
his/her extremal branch after 100 years of lifetime would be 
$$
    2^{-10^{12}}
$$
while a tunnel amplitude with objects in the range of SI units%
\footnote{The extreme numerics of macroscopic tunnel amplitudes motivated this author's  
previous approach to objectification \cite{Polley2005} in which tunnel amplitudes where 
considered as the fluctuating quantity. However, while a particular summand in a superposition 
of object-observer states might well be strongly enhanced this way, alternative branches 
are still there, and with a chance of about 50\%, the next measurement will enhance a 
previously suppressed branch by an even larger factor.
A series of determined outcomes can only emerge if an observer's histories are 
finite, and if one of them is selected as an entity.} would be
$$
    e^{-10^{34}}
$$
So there is some tolerance for neglecting the remote possibilities. 
Smallness, in itself, need not be a problem, since awareness requires interaction (as assumed
throughout the paper) and there is no handle for an interaction with the amplitude in 
strictly linear dynamics. However, what the scenario suggests 
is that $2^{-10^{12}}$ is big enough for an observer to be aware of, while 
$e^{-10^{34}}$ is not. Discretisation of amplitudes could be a solution, but
it would have to occur on a scale vastly different from what is found ($10^{-20}$) 
by considering fundamental limits on the amount of information needed to {\em measure\/}
amplitudes by using interference patterns \cite{Zee2006,Mueller2009}. 

The role assumed for an observer's consciousness allows to avoid a conceptional dilemma. 
If it were the number of {\em all} physical branchings that would distinguish the extremal 
branch, observer's awareness would reside in precisely those parts of his/her wavefunction 
whose amplitudes are the smallest among alternatives. However, the extremal branch is only 
distinguished by the number of subsystems activated for cognitive processing. This would 
seem to be compatible with an equal number of physical (and mostly unperceived) branchings, 
more or less, in each of an observer's histories.

\section{Conclusion\label{Conclusion}} 

Assuming an observer's wavefunction to be branching, by unitary evolution,
into a large number of decoherent parts, and assuming each bit of observer's 
experience to be physically represented by a subsystem, it is conjectured that
the number of subsystems {\em cognitively processed\/} near the ends of branches be 
stochastically independent and power-law distributed. 
Order statistics then implies that one of the branches carries almost all 
of the dimension contained in the ensemble of those subsystems. 
This would render physical substance to an observer's impression of living in a unique 
world. 

The storing of records in interactive subsystems ensures objectivity 
to the extent that an observer is able to verify it. 
On his/her extremal branch, he/she will find other observers interacting with the same
subsystem, so they will agree on what has been recorded. 

As to whether quantum physics is determinate or indeterminate, the scenario 
suggests a complex answer. The law governing all time evolution is a 
Schr\"odinger equation, deterministic in terms of an all-encompassing wavefunction. 
What is perceived to happen at a particular moment in an observer's lifetime, however, is 
determined by an effective ``final'' Hamiltonian, leaving only probabilistic ways of 
predicting anything at intermediate times. In principle, the effective ``final'' Hamiltonian 
should be determined by the all-encompassing wavefunction at any time, in conjunction with a
universal exact Hamiltonian. 

The scenario suggests a fundamental role of backward causation in quan\-tum-mech\-anical 
objectification.  How plausible is it that conscious perception of the present should 
not only depend on existing memories, as in the biological theory of \cite{Edelman1989}, 
but on the processing of memories in the future? 
The Heisenberg picture may provide a clue. 
State vectors in this picture are
``timeless'', but time reappears in ``labels'' of basis states. The state 
$|\psi\rangle$ of an observer's lifetime, including all of his/her experiences, can be 
expressed in many different bases, like  
$|\psi\rangle = \sum_{l_1} \psi_{l_1,t_1} \, |l_1,t_1\rangle 
              = \sum_{l_2} \psi_{l_2,t_2} \, |l_2,t_2\rangle$ etc.
The number of basis states needed to represent $|\psi\rangle$ is exhaustively given  
by the dimension of the extremal branch. But at intermediate times, the observer's state 
is the same, only expressed in different bases. The question is thus shifted
from causal relations to what makes up the time label of basis states.

\end{document}